\documentclass[superscriptaddress,twocolumn,amsmath,amssymb,footinbib]{revtex4-1}
\usepackage[english]{babel}
\usepackage[pdftex]{graphicx}
\usepackage{gensymb}
\usepackage{xcolor}
\usepackage{soul}
\usepackage{mathrsfs}
\begin{document}

\newcommand{\physicsDept}{Department of Physics, New Mexico State University, Las Cruces, NM 88001, USA}
\newcommand{\CHMEDept}{Department of Chemical and Materials Engineering, New Mexico State University, Las Cruces, NM 88001, USA}
\newcommand{\ang}{\text{\normalfont\AA}}
\newcommand{\dpar}{$d_{||}$ }
\newcommand{\dxzyz}{$(d_{\text{xz}}$, $d_{\text{yz}})$ }
\newcommand{\red}{\textcolor{red}}

\title{Monopole Traps for Position-Based Information Coding}

\author{Prakash Timsina}
\affiliation{\physicsDept}
\author{Andres Chappa}
\affiliation{\physicsDept}
\author{Deema Alyones}
\affiliation{\CHMEDept}
\affiliation{\physicsDept}
\author{Boris Kiefer}
\affiliation{\physicsDept}
\author{Ludi Miao*}
\affiliation{\physicsDept}

\date{\today}


\begin{abstract}
We propose a spin-ice-based heterostructure capable of encoding magnetic monopole quasiparticle positions for non-volatile information storage applications. Building upon two-dimensional magnetic monopole gases formed at the interface between 2-in–2-out spin ice and all-in–all-out antiferromagnetic pyrochlore iridate, the design introduces a 3-in-1-out/1-in-3-out fragmented barrier layer into the spin-ice matrix, defining two energetically stable monopole traps. The occupancy of these traps can be deterministically controlled by an externally applied magnetic field. Monte Carlo simulations reveal robust bistable switching, thermal stability below 0.22 K, and fully reversible field-driven transitions, demonstrating the system’s potential for reliable, repeatable memory operation. Crucially, the heterostructure exhibits emergent ferromagnetism linked to monopole position, enabling non-destructive readout of the memory state via spatially resolved magnetic imaging. Unlike topological carriers such as skyrmions, monopoles confined at the sub-nanometer scale offer three orders of magnitude higher information density. These results establish these monopole-trap heterostructures as a scalable platform for next-generation ultra-compact memory technologies.
\end{abstract}

\maketitle

\section{Introduction}
    Magnetic monopoles, which emerge as quasiparticle excitations in spin-ice systems, have attracted significant attention due to their unconventional properties that transcend Maxwell’s equations and their potential for transformative applications in information technology \cite{Moessner2003,Castelnovo2008,Fennell2009Science,Bramwell2009,Ladak2010,Mengotti2011,Sala2012,Farhan2013,Pan2016,schiffer2021artificial}. To date, most studies have focused on bulk spin-ice materials \cite{Harris1997,fennell2005neutron,jaubert2011magnetic,Fennell2007,Pan2016,Morris2009,sarkar2014dynamics,Edberg2025,Fennell2007,Bramwell2001,snyder2004low,powell2025dynamic,Dusad2019}
, where monopoles and antimonopoles appear in equal numbers, resulting in a zero net magnetic charge and a ground state devoid of monopoles \cite{Castelnovo2008,Bramwell2009}. This intrinsic charge neutrality constrains the manipulation of monopoles in ways analogous to electrons in conventional electronic systems, thereby limiting their utility for practical device applications. In contrast, our recent work introduces a two-dimensional magnetic monopole gas (2DMG), characterized by unequal monopole populations and even isolated single charges \cite{Miao2020,Timsina2024}. This 2DMG is tunable via external magnetic fields, offering a promising platform for controllable monopole dynamics and future device applications.

Building upon these advancements, here we propose a spin-ice-based oxide heterostructure capable of trapping magnetic monopoles in one of two spatially separated spin-ice regions, thereby enabling non-volatile, position-based information encoding. The heterostructure $R_2$Ir$_2$O$_7$/$R_2$Ti$_2$O$_7$/$R'_2$Ir$_2$O$_7$/$R_2$Ti$_2$O$_7$/$R_2$Ir$_2$O$_7$ ($R,R'$ are rare earth elements) consists of two-in-two-out (2I2O) spin-ice/all-in-all-out (AIAO) antiferromagnetic interfaces $R_2$Ti$_2$O$_7$/$R_2$Ir$_2$O$_7$ which host 2DMGs \cite{Miao2020,Timsina2024}, inserted by a central magnetic barrier layer $R'_2$Ir$_2$O$_7$ that exhibits a one-in-three-out(1I3O)/three-in-one-out(3I1O) fragmented spin configuration. This barrier divides the spin-ice region into two well-defined monopole traps and suppresses monopole diffusion across the barrier. By applying a magnetic field during cooling, monopoles can be steered into either trap depending on the field direction. After the field is removed, the fragmented barrier preserves the trapped configuration, thereby encoding the cooling history in the monopole position. We also demonstrate that the monopole distribution exhibits clear bistability and robust thermal stability below a characteristic leakage temperature $T_\text{leak}=0.22$ K. The monopole state can also be isothermally switched back and forth between traps through magnetic field cycles, with high fidelity. The switching process is robust and reproducible across many cycles, confirming the system’s reliability as a memory element. We further show that the system develops emergent ferromagnetism, where the polarization reflects the locations of monopoles, providing a means to read out their positions. Our findings establish a practical mechanism for controlling and stabilizing magnetic monopoles in solid-state systems, laying the groundwork for their application in future information technologies.

\section{Results and discussions}
In pyrochlore iridate $R_2$Ir$_2$O$_7$, the relative strength of the $d$–$f$ interaction, expressed as $H_\text{loc}$/$J_\text{eff}$ between Ir$^\text{4+}$ and $R^\text{3+}$, gives rise to various spin structures \cite{Lefrancois2017,timsina2025emergent,Guo2016}: 2I2O for $0 < H_\text{loc}$/$J_\text{eff} < 2$; 1I3O/3I1O fragmented phases for $2 < H_\text{loc}$/$J_\text{eff} < 6$; and an antiferromagnetic AIAO phase for $H_\text{loc}$/$J_\text{eff} > 6$. The strength of this $d$–$f$ interaction, and thus the resulting magnetic structure, can be tuned by selecting the appropriate rare-earth ion. For example, Ho$_2$Ir$_2$O$_7$ exhibits a fragmented phase \cite{Lefrancois2017}, while Nd$_2$Ir$_2$O$_7$ stabilizes an AIAO structure \cite{Guo2016}. 
In $R_2$Ir$_2$O$_7$/$R_2$Ti$_2$O$_7$/$R_2$Ir$_2$O$_7$ heterostructures, the interface between the antiferromagnetic $R_2$Ir$_2$O$_7$ layer and the spin-ice $R_2$Ti$_2$O$_7$ layer imposes a boundary condition that prevents the ice rule from being satisfied at every tetrahedral site in the spin-ice region. As a result, a 2DMG with a non-zero net magnetic charge in the spin-ice layer becomes inevitable \cite{Miao2020}. However, the position and distribution of these monopoles are governed solely by the system’s the system’s free energy landscape, and thus cannot be directly used to encode nonvolatile information.

\subsection{The Role of the Fragmented Barrier}
\begin{figure*}
\begin{center}
\includegraphics[width=7in]{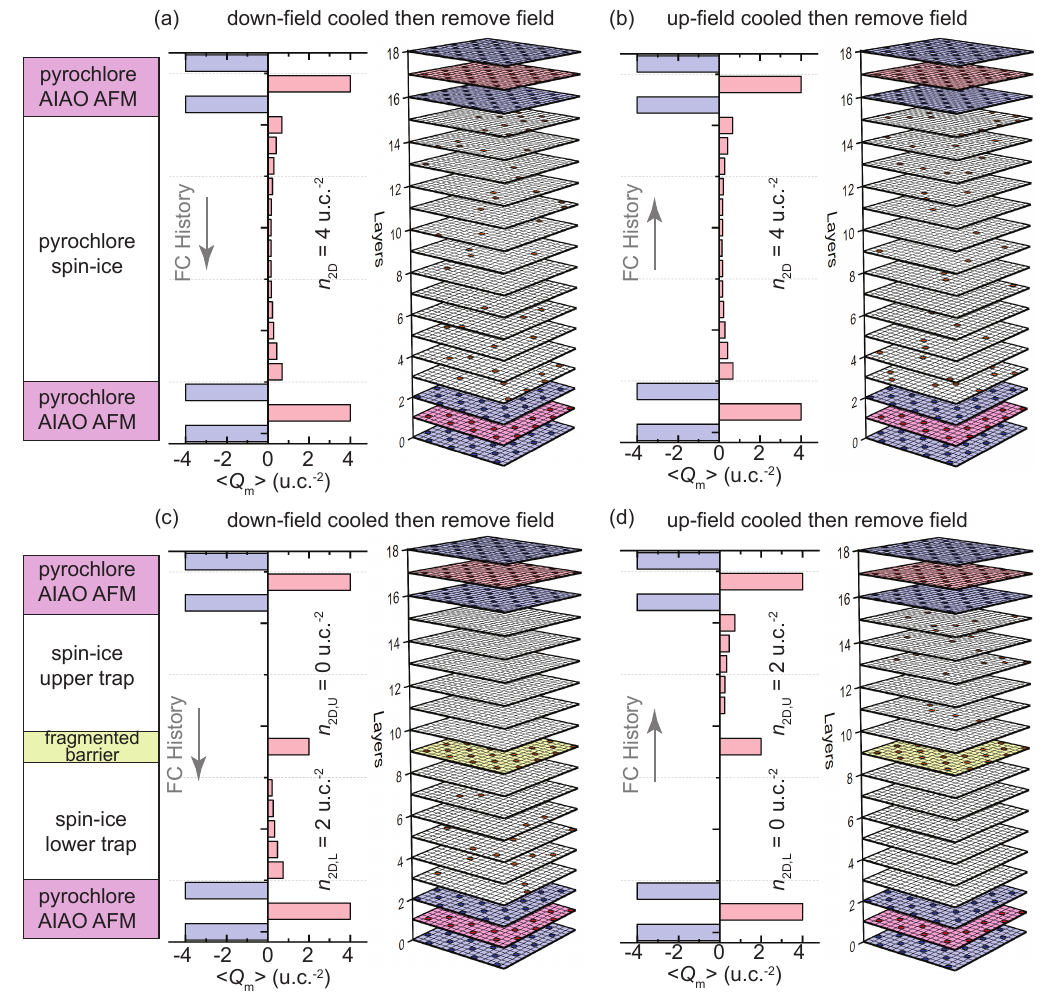}
\end{center}
\caption{Monte-Carlo-simulated averaged monopole profiles and a single snapshot of monopole distribution of a $R_2$Ir$_2$O$_7$/$R_2$Ti$_2$O$_7$/$R_2$Ir$_2$O$_7$ heterostructure, after the heterostructure is magnetic-field-cooled with (a) field pointing down and (b) field pointing up to ground state and then remove the field. Monte-Carlo-simulated averaged monopole profiles and a single snapshot of monopole distribution of a $R_2$Ir$_2$O$_7$/$R_2$Ti$_2$O$_7$/$R'_2$Ir$_2$O$_7$/$R_2$Ti$_2$O$_7$/$R_2$Ir$_2$O$_7$ heterostructure with fragmented pyrochlore barrier, after the heterostructure is magnetic-field-cooled with (c) field pointing down and (d) field pointing up to ground state and then remove the field. $H_\text{loc}$/$J_\text{eff}$ is set at 14 for $R_2$Ir$_2$O$_7$ to create 2DMG in the heterostructure and $H'_\text{loc}$/$J_\text{eff}$ is set at 4 for $R'_2$Ir$_2$O$_7$ to serve as monopole barrier layer. The insets are illustration of the heterostructures.}
\end{figure*}

To demonstrate this point, we use the Monte Carlo simulation to investigate the spin structure of an $R_2$Ir$_2$O$_7$/$R_2$Ti$_2$O$_7$/$R_2$Ir$_2$O$_7$ heterostructure \cite {Miao2020,Timsina2024}. For simplicity, we pick a (001)-oriented heterostructure, with $H_\text{loc}$/$J_\text{eff}$ = 14 (antiferromagnetic) for the top and bottom layers to enable 2DMG in the spin-ice layer, and only consider nearest-neighbor interactions. Therefore the Hamiltonian is:
\begin{equation}
\label{Eq:twoadded}
\mathscr{H} = J_{\text{eff}} \sum_{\langle i,j \rangle} \sigma_i \sigma_j + \frac{1}{6} H_{\text{loc}} \sum_{\langle i,\alpha \rangle} \sigma_i \sigma_{\alpha}
\end{equation}

where $\sigma_i$, and  $\sigma_j$ (both taking values $\pm$1) represent the Ising pseudo-spins of $R^{3+}$ and $\sigma_{\alpha}$ (taking values $\pm$1) indicates the Ising pseudo-spins of Ir$^{4+}$, oriented towards or away from a tetrahedron. The notation $\langle i, j \rangle$ denotes summation over nearest-neighbor sites. The parameter $J_{\text{eff}}$ represents the effective nearest-neighbor interaction between $R^{3+}$ moments. And, $H_{\text{loc}}$ is a static local background field generated by the Ir$^{4+}$ moments over the $R^{3+}$ moments.

In our Monte Carlo simulations, we applied magnetic fields perpendicular to the heterostructure interface during simulated annealing—cooling from high temperature toward the ground state. We demonstrate two cases, where the cooling field points downward and upward, respectively. With the field present, the monopole distribution responds asymmetrically, accumulating on opposite sides of the heterostructure depending on the field direction (see supplementary information Fig. S1). However, after reaching the ground state and removing the field, the monopole distribution in both cases relaxes to the same equilibrium configuration, as shown in Fig. 1(a) and 1(b). This final state is governed solely by spin-ice and monopole entropies \cite{Timsina2024}, indicating that the field-cooling history is not maintained when the barrier layer is absent.

To prevent the 2DMG from rebalancing after magnetic field removal, we inserted a fragmented pyrochlore iridate layer, $R'_2$Ir$_2$O$_7$, into the center of the spin-ice layer, as illustrated in Fig. 1(c). This barrier layer, with $H'_\text{loc}/J_\text{eff}$ tentatively set at 4, stabilizes a 3I1O/1I3O fragmented spin state, which creates both energetic and entropic resistance to monopole transport, dividing the spin-ice into two separate traps.
We applied a magnetic field pointing downward during cooling, biasing monopoles toward the lower trap. After reaching the ground state, we removed the field and allowed the system to relax. Unlike the case without the barrier, where the monopoles redistribute symmetrically, here, the fragmented layer pins the monopoles in the lower trap, preserving the memory of the field direction, as shown in Fig. 1(c). This is evident by the resulting monopole density of $n_{2D,\text{up}} = 0$ u.c.$^{-2}$ in the upper trap and $n_{2D,\text{low}} = 2$ u.c.$^{-2}$ in the lower trap. To test reversibility, we repeated the simulation with the cooling field pointing upward. In this case, the monopoles accumulated in the upper trap and remained localized there after the field was removed, as shown in Fig. 1(d). These results demonstrate that the fragmented layer traps monopoles in one of two stable configurations, depending on the polarity of the cooling field history, thus enabling bistable and non-volatile information storage.

\subsection{Thermal Stability of Monopole Switching}

\begin{figure*}
\begin{center}
\includegraphics[width=7in]{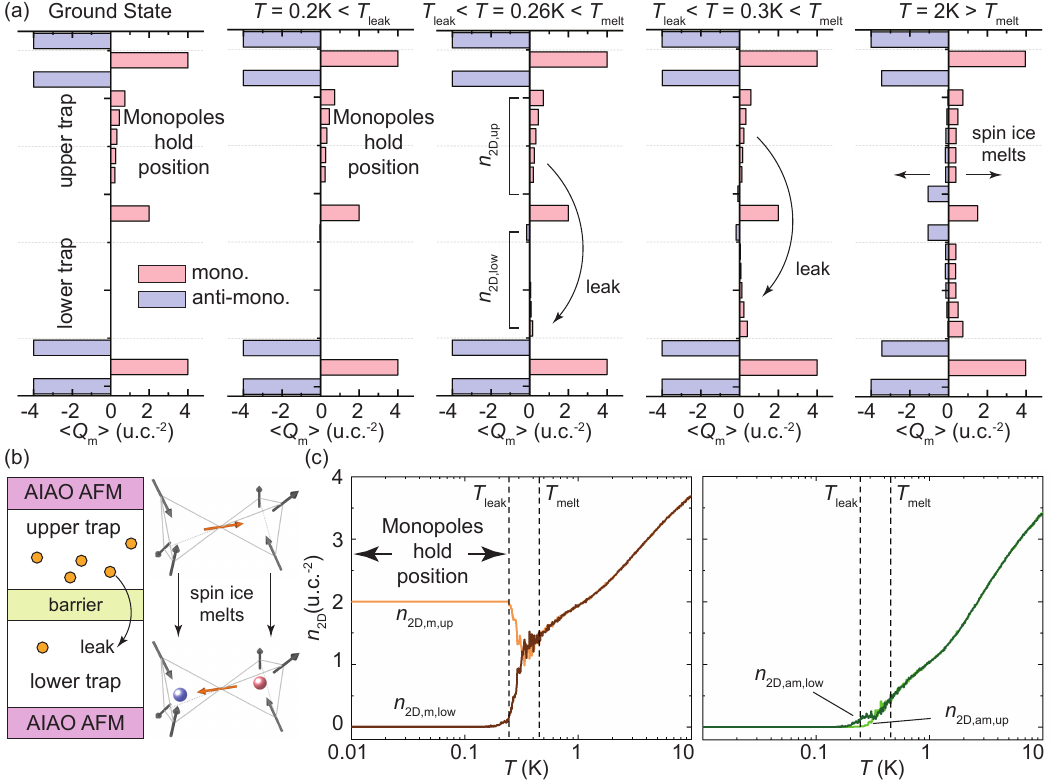}
\end{center}
\caption{(a) Monopole profiles of a field-cooled and then zero-field-warmed up $R_2$Ir$_2$O$_7$/$R_2$Ti$_2$O$_7$/$R'_2$Ir$_2$O$_7$/$R_2$Ti$_2$O$_7$/$R_2$Ir$_2$O$_7$ heterostructure, where $T_\text{leak}\sim 0.22$ K is the temperature where monopole start to leave the upper trap and enter the lower trap, and $T_\text{melt}\sim 0.45$ K is temperature where monopole–antimonopole pairs are thermally created in the spin-ice layer due to the break down of the ice rule. (b) Illustration of the monopole leak and spin-ice melt process. (c) 2D densities of monopoles (left) and antimonopoles (right) in the upper trap and lower trap as a function of zero-field warm-up temperature.}
\end{figure*}

To evaluate the thermal stability of the field-cooled monopole states, we performed zero-field warm-up simulations, starting from the bistable configurations shown in Figs. 1(c) and 1(d). In these simulations, the system is gradually heated while the external magnetic field remains off, and monopole distribution profiles are monitored as functions of temperature.

As shown in Fig. 2(a), monopoles remain well confined in their respective traps at low temperatures. For example, when initialized with monopoles in the upper trap, the density $n_{2D,\text{up}}$ remains constant, while $n_{2D,\text{low}}$ stays negligible up to approximately $0.26$ K. Beyond this point, monopoles begin to thermally overcome the central fragmented barrier and slowly diffuse into the lower trap, resulting in a gradual redistribution of magnetic charge, until the upper and lower traps reach a similar monopole density at $\sim$0.3 K. As the temperature continues to rise, a second critical threshold is reached at $T_\text{melt} \sim 0.45$ K, where the spin-ice rules begin to break down and thermally activated monopole–antimonopole pairs emerge throughout the spin-ice layer. This marks the transition from a memory-retaining regime to a disordered, entropically dominated monopole plasma. These two-stage processes—monopole leakage followed by spin-ice melting—are schematically illustrated in Fig. 2(b).

Figure 2(c) quantifies this evolution by tracking the 2D densities of monopoles and antimonopoles in both traps as a function of temperature. Below $T_\text{leak} \sim 0.22$ K, the system maintains its initial monopole configuration. Between $T_\text{leak}$ and $T_\text{melt} \sim 0.45$ K, monopoles begin to leak across the barrier, while pair creation remains absent. Above $T_\text{melt}$, both traps exhibit increasing populations of oppositely charged monopoles, signaling the onset of spin-ice melting.

\subsection{Magnetic Field-Controlled Switching and Emergent Ferromagnetism}

\begin{figure*}
\begin{center}
\includegraphics[width=7in]{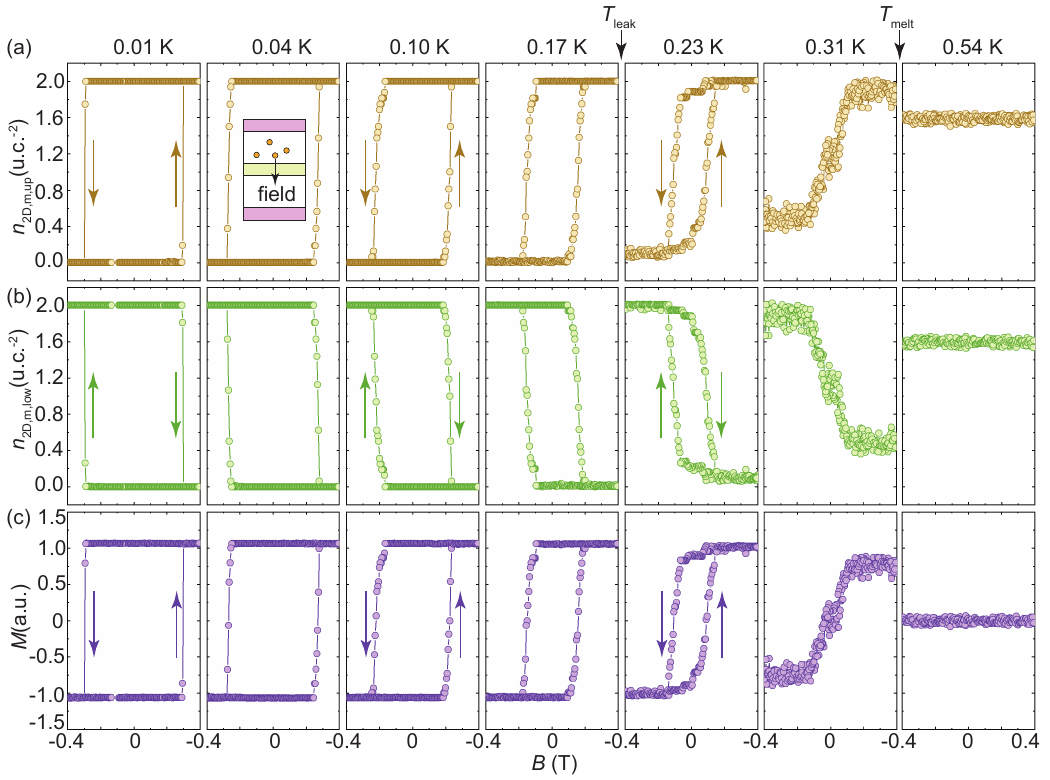}
\end{center}
\caption{2D densities of monopoles in (a) the upper trap, (b) lower trap and (c) net magnetization as a function of magnetic field, as 
 field is sweeping between -0.4 T to 0.4 T at various fixed temperatures. Field scanning directions are labeled by arrows.}
\end{figure*}
Next, we investigated whether the position of trapped monopoles could be reversibly manipulated using an external magnetic field. To this end, we performed field-sweeping simulations at several fixed temperatures, starting from a zero-field-cooled state in which monopoles were initially localized in one of the two traps. A magnetic field perpendicular to the heterostructure was continuously swept from $-0.4$ T to $+0.4$ T and back, while the 2D monopole densities in the upper and lower traps were tracked throughout the process.

Figure 3 (a) and (b) presents these densities as functions of magnetic field at temperatures ranging from 0.01 K to 0.54 K. At low temperatures (e.g., 0.01 K), we observe sharp and hysteretic switching: as the field crosses a threshold, monopoles abruptly relocate from one trap to the other, indicating robust bistability and a well-defined coercive field. The densities in the two traps are strongly anti-correlated, indicating the field-driven-monopole-transfer across the barrier. As the temperature increases toward $T_\text{leak}$, the hysteresis loop narrows, implying thermal assistance in overcoming the barrier. For $T_\text{leak} < T < T_\text{melt}$, the fidelity of the switching begins to degrade as temperature increases, as thermal fluctuations partially randomize the monopole distribution and reduce the contrast between the two states. Near $T_\text{melt}$, switching becomes smoother and nearly reversible, while above $T_\text{melt}$ (e.g., 0.54 K), no discernible switching or hysteresis is observed, consistent with a thermally disordered monopole plasma. To assess reproducibility, we simulated repeated magnetic field cycles below $T_\text{leak}$, confirming that the monopole-trap heterostructure exhibits clean, drift-free, and fully reversible switching.

Having established field-controlled, reversible monopole switching, we now address the critical requirement of reading the information state. The net magnetization of the heterostructure serves as a readout signal, emerging from the asymmetric spin configurations induced by monopole positioning in the frustrated spin-ice layer. Although none of the constituent layers—2I2O, 3I1O/1I3O, or AIAO—is intrinsically ferromagnetic, the full heterostructure exhibits an emergent ferromagnetic response below $T_\text{leak}$, as evidenced by the distinct magnetic moments closely associated with monopoles trapped in the upper versus lower region [as shown in Fig. 3(c)]. This bistable magnetic moment, originating from bistable monopole positions, enables non-destructive readout using local probes such as scanning Superconducting Quantum Interference Device (SQUID) microscopy, offering a practical route for reading monopole-based memory states.

\subsection{Discussion}

Our simulations focused on nearest-neighbor interactions, but incorporating long-range dipolar forces—present in real spin-ice systems—would likely enhance monopole repulsion across traps \cite{Miao2020}, further stabilizing bistable states and sharpening switching transitions. We also tested various lateral system sizes and found that monopole trapping, thermal stability, and reversible switching remain robust, confirming these effects are not finite-size artifacts. While this study used (001)-oriented heterostructures, which require relatively high $H_\text{loc}/J_\text{eff}>12$ in $R_2$Ir$_2$O$_7$ to generate a monopole gas, our previous work shows that switching to a (111) orientation significantly lowers this threshold to $H_\text{loc}/J_\text{eff}>3$, offering greater flexibility in material and device design.

While this work demonstrates a binary memory scheme using two monopole traps, the concept can be extended to higher information densities and more complex logic. For example, in a multilayer heterostructure design such as antiferromagnetic/spin-ice/barrier/spin-ice/barrier/spin-ice/antiferromagnetic, additional fragmented barriers divide the spin-ice region into multiple isolated trapping zones. This architecture would allow monopoles to be selectively positioned in one of three or more stable locations, effectively enabling multi-level memory encoding and supporting logic operations beyond the conventional binary state.

The monopole trap operates at ultra-low temperatures (below 0.2 K), comparable to those used in quantum computing platforms. These conditions ensure high fidelity in monopole localization and memory retention. At the same time, the concept is not limited to such regimes; it can be extended to systems with stronger interactions and higher characteristic energy scales—such as high-temperature spin-ice candidates in spinel iridates (e.g., IrO$_2$ \cite{Onoda2019}) or artificial spin-ice arrays \cite{Ladak2010}—potentially enabling room-temperature operation. Moreover, temperature introduces a trade-off in performance: lower temperatures enhance the precision and stability of the monopole’s positional state, while higher temperatures reduce the magnetic field required for writing, suggesting a tunable balance point for optimal device operation.

In comparison with topological carriers such as skyrmions \cite{nagaosa2013topological,fert2017magnetic,li2023magnetic}—which require lateral sizes on the order of tens of nanometers for stability—the monopole states in our heterostructure can be confined to sub-nanometer dimensions. This scale difference translates directly into storage density: while skyrmion-based memories typically achieve 1 bit per 1000 nm$^2$, monopole-trap devices could support densities three orders of magnitude higher, approaching 1 bit per 0.5 nm$^2$.

If extended into the quantum regime—such as in quantum spin-ice systems \cite{Pan2014,Pan2016}—the monopole trap architecture could enable quantum information encoding. Monopoles in coherent superpositions of positions across the two traps would encode spatial qubits. Since monopole position directly determines net magnetization, and total magnetization reflects conserved angular momentum, the positions of two monopoles can become entangled as a consequence of angular momentum conservation. This quantum extension of the 2DMG could support non-local entanglement and coherent quantum control, offering a pathway toward quantum memory and fault-tolerant computation.

\section{Conclusion}

In this work, we proposed and simulated a spin-ice-based heterostructure that traps magnetic monopoles in bistable spatial configurations, enabling non-volatile information storage based on monopole positions. By embedding a fragmented pyrochlore barrier within the spin-ice layer, we demonstrated localization of monopoles in either trap can be realized by a training field during cooling, with the monopole state preserved after field removal. Monte Carlo simulations revealed robust, repeatable switching behavior under magnetic field cycling and thermal resilience across a relevant operating range. We demonstrated that the system exhibits emergent ferromagnetism with polarization tied to monopole positions, enabling monopole position readout. We also discussed the temperature dependence of switching dynamics, scalability of information density, and prospects for quantum encoding. These results open a new avenue for exploring nonequilibrium dynamics in magnetic monopole systems, while also serving as a promising platform for compact and scalable information technologies.

\section{Methods}
The spin model employed in this work is based on Monte Carlo simulations designed to explore the accessible spin configurations of the proposed monopole-trap heterostructure and determine the thermodynamically favorable spin states. The simulations incorporate nearest-neighbor exchange interactions among Ising spins. A single spin-flip Metropolis algorithm \cite{Lefrancois2017} with periodic boundary conditions is used, where spins are randomly flipped and accepted or rejected according to Boltzmann statistics. The simulations are performed on $8 \times 8 \times 8$ lattice sites, representing the full $R_2$Ir$_2$O$_7$/$R_2$Ti$_2$O$_7$/$R'_2$Ir$_2$O$_7$/$R_2$Ti$_2$O$_7$/$R_2$Ir$_2$O$_7$ heterostructure. For each run, approximately $10^{5}$ single spin-flip attempts are made following 1000 thermalization steps to ensure equilibrium. From the resulting spin configurations, we extract monopole distributions by evaluating the net magnetic charge within each tetrahedron. This charge mapping allows us to resolve spatial monopole profiles across the heterostructure layers. Additional details of the Monte Carlo implementation, including data structures and optimization strategies, are documented in prior work \cite{Miao2020,Timsina2024}. The code was developed in C++ specifically for the heterostructure geometry considered in this study.

\section*{Data availability}
All relevant data are available from the authors upon reasonable requests.

\section*{Code availability}
All relevant codes are available from the authors upon reasonable requests.

\bibliography{references}

\begin{thebibliography}{31}%
\makeatletter
\providecommand \@ifxundefined [1]{%
 \@ifx{#1\undefined}
}%
\providecommand \@ifnum [1]{%
 \ifnum #1\expandafter \@firstoftwo
 \else \expandafter \@secondoftwo
 \fi
}%
\providecommand \@ifx [1]{%
 \ifx #1\expandafter \@firstoftwo
 \else \expandafter \@secondoftwo
 \fi
}%
\providecommand \natexlab [1]{#1}%
\providecommand \enquote  [1]{``#1''}%
\providecommand \bibnamefont  [1]{#1}%
\providecommand \bibfnamefont [1]{#1}%
\providecommand \citenamefont [1]{#1}%
\providecommand \href@noop [0]{\@secondoftwo}%
\providecommand \href [0]{\begingroup \@sanitize@url \@href}%
\providecommand \@href[1]{\@@startlink{#1}\@@href}%
\providecommand \@@href[1]{\endgroup#1\@@endlink}%
\providecommand \@sanitize@url [0]{\catcode `\\12\catcode `\$12\catcode `\&12\catcode `\#12\catcode `\^12\catcode `\_12\catcode `\%12\relax}%
\providecommand \@@startlink[1]{}%
\providecommand \@@endlink[0]{}%
\providecommand \url  [0]{\begingroup\@sanitize@url \@url }%
\providecommand \@url [1]{\endgroup\@href {#1}{\urlprefix }}%
\providecommand \urlprefix  [0]{URL }%
\providecommand \Eprint [0]{\href }%
\providecommand \doibase [0]{http://dx.doi.org/}%
\providecommand \selectlanguage [0]{\@gobble}%
\providecommand \bibinfo  [0]{\@secondoftwo}%
\providecommand \bibfield  [0]{\@secondoftwo}%
\providecommand \translation [1]{[#1]}%
\providecommand \BibitemOpen [0]{}%
\providecommand \bibitemStop [0]{}%
\providecommand \bibitemNoStop [0]{.\EOS\space}%
\providecommand \EOS [0]{\spacefactor3000\relax}%
\providecommand \BibitemShut  [1]{\csname bibitem#1\endcsname}%
\let\auto@bib@innerbib\@empty
\bibitem [{\citenamefont {Moessner}\ and\ \citenamefont {Sondhi}(2003)}]{Moessner2003}%
  \BibitemOpen
  \bibfield  {author} {\bibinfo {author} {\bibfnamefont {R.}~\bibnamefont {Moessner}}\ and\ \bibinfo {author} {\bibfnamefont {S.~L.}\ \bibnamefont {Sondhi}},\ }\href@noop {} {\bibfield  {journal} {\bibinfo  {journal} {Physical Review B}\ }\textbf {\bibinfo {volume} {68}},\ \bibinfo {pages} {064411} (\bibinfo {year} {2003})}\BibitemShut {NoStop}%
\bibitem [{\citenamefont {Castelnovo}\ \emph {et~al.}(2008)\citenamefont {Castelnovo}, \citenamefont {Moessner},\ and\ \citenamefont {Sondhi}}]{Castelnovo2008}%
  \BibitemOpen
  \bibfield  {author} {\bibinfo {author} {\bibfnamefont {C.}~\bibnamefont {Castelnovo}}, \bibinfo {author} {\bibfnamefont {R.}~\bibnamefont {Moessner}}, \ and\ \bibinfo {author} {\bibfnamefont {S.~L.}\ \bibnamefont {Sondhi}},\ }\href@noop {} {\bibfield  {journal} {\bibinfo  {journal} {Nature}\ }\textbf {\bibinfo {volume} {451}},\ \bibinfo {pages} {42} (\bibinfo {year} {2008})}\BibitemShut {NoStop}%
\bibitem [{\citenamefont {Fennell}\ \emph {et~al.}(2009)\citenamefont {Fennell}, \citenamefont {Deen}, \citenamefont {Wildes}, \citenamefont {Schmalzl}, \citenamefont {Prabhakaran}, \citenamefont {Boothroyd}, \citenamefont {Aldus}, \citenamefont {McMorrow},\ and\ \citenamefont {Bramwell}}]{Fennell2009Science}%
  \BibitemOpen
  \bibfield  {author} {\bibinfo {author} {\bibfnamefont {T.}~\bibnamefont {Fennell}}, \bibinfo {author} {\bibfnamefont {P.}~\bibnamefont {Deen}}, \bibinfo {author} {\bibfnamefont {A.}~\bibnamefont {Wildes}}, \bibinfo {author} {\bibfnamefont {K.}~\bibnamefont {Schmalzl}}, \bibinfo {author} {\bibfnamefont {D.}~\bibnamefont {Prabhakaran}}, \bibinfo {author} {\bibfnamefont {A.}~\bibnamefont {Boothroyd}}, \bibinfo {author} {\bibfnamefont {R.}~\bibnamefont {Aldus}}, \bibinfo {author} {\bibfnamefont {D.}~\bibnamefont {McMorrow}}, \ and\ \bibinfo {author} {\bibfnamefont {S.}~\bibnamefont {Bramwell}},\ }\href@noop {} {\bibfield  {journal} {\bibinfo  {journal} {Science}\ }\textbf {\bibinfo {volume} {326}},\ \bibinfo {pages} {415} (\bibinfo {year} {2009})}\BibitemShut {NoStop}%
\bibitem [{\citenamefont {Bramwell}\ \emph {et~al.}(2009)\citenamefont {Bramwell}, \citenamefont {Giblin}, \citenamefont {Calder}, \citenamefont {Aldus}, \citenamefont {Prabhakaran},\ and\ \citenamefont {Fennell}}]{Bramwell2009}%
  \BibitemOpen
  \bibfield  {author} {\bibinfo {author} {\bibfnamefont {S.~T.}\ \bibnamefont {Bramwell}}, \bibinfo {author} {\bibfnamefont {S.}~\bibnamefont {Giblin}}, \bibinfo {author} {\bibfnamefont {S.}~\bibnamefont {Calder}}, \bibinfo {author} {\bibfnamefont {R.}~\bibnamefont {Aldus}}, \bibinfo {author} {\bibfnamefont {D.}~\bibnamefont {Prabhakaran}}, \ and\ \bibinfo {author} {\bibfnamefont {T.}~\bibnamefont {Fennell}},\ }\href@noop {} {\bibfield  {journal} {\bibinfo  {journal} {Nature}\ }\textbf {\bibinfo {volume} {461}},\ \bibinfo {pages} {956} (\bibinfo {year} {2009})}\BibitemShut {NoStop}%
\bibitem [{\citenamefont {Ladak}\ \emph {et~al.}(2010)\citenamefont {Ladak}, \citenamefont {Read}, \citenamefont {Perkins}, \citenamefont {Cohen},\ and\ \citenamefont {Branford}}]{Ladak2010}%
  \BibitemOpen
  \bibfield  {author} {\bibinfo {author} {\bibfnamefont {S.}~\bibnamefont {Ladak}}, \bibinfo {author} {\bibfnamefont {D.}~\bibnamefont {Read}}, \bibinfo {author} {\bibfnamefont {G.}~\bibnamefont {Perkins}}, \bibinfo {author} {\bibfnamefont {L.}~\bibnamefont {Cohen}}, \ and\ \bibinfo {author} {\bibfnamefont {W.}~\bibnamefont {Branford}},\ }\href@noop {} {\bibfield  {journal} {\bibinfo  {journal} {Nature Physics}\ }\textbf {\bibinfo {volume} {6}},\ \bibinfo {pages} {359} (\bibinfo {year} {2010})}\BibitemShut {NoStop}%
\bibitem [{\citenamefont {Mengotti}\ \emph {et~al.}(2011)\citenamefont {Mengotti}, \citenamefont {Heyderman}, \citenamefont {Rodr{\'\i}guez}, \citenamefont {Nolting}, \citenamefont {H{\"u}gli},\ and\ \citenamefont {Braun}}]{Mengotti2011}%
  \BibitemOpen
  \bibfield  {author} {\bibinfo {author} {\bibfnamefont {E.}~\bibnamefont {Mengotti}}, \bibinfo {author} {\bibfnamefont {L.~J.}\ \bibnamefont {Heyderman}}, \bibinfo {author} {\bibfnamefont {A.~F.}\ \bibnamefont {Rodr{\'\i}guez}}, \bibinfo {author} {\bibfnamefont {F.}~\bibnamefont {Nolting}}, \bibinfo {author} {\bibfnamefont {R.~V.}\ \bibnamefont {H{\"u}gli}}, \ and\ \bibinfo {author} {\bibfnamefont {H.-B.}\ \bibnamefont {Braun}},\ }\href@noop {} {\bibfield  {journal} {\bibinfo  {journal} {Nature Physics}\ }\textbf {\bibinfo {volume} {7}},\ \bibinfo {pages} {68} (\bibinfo {year} {2011})}\BibitemShut {NoStop}%
\bibitem [{\citenamefont {Sala}\ \emph {et~al.}(2012)\citenamefont {Sala}, \citenamefont {Castelnovo}, \citenamefont {Moessner}, \citenamefont {Sondhi}, \citenamefont {Kitagawa}, \citenamefont {Takigawa}, \citenamefont {Higashinaka},\ and\ \citenamefont {Maeno}}]{Sala2012}%
  \BibitemOpen
  \bibfield  {author} {\bibinfo {author} {\bibfnamefont {G.}~\bibnamefont {Sala}}, \bibinfo {author} {\bibfnamefont {C.}~\bibnamefont {Castelnovo}}, \bibinfo {author} {\bibfnamefont {R.}~\bibnamefont {Moessner}}, \bibinfo {author} {\bibfnamefont {S.~L.}\ \bibnamefont {Sondhi}}, \bibinfo {author} {\bibfnamefont {K.}~\bibnamefont {Kitagawa}}, \bibinfo {author} {\bibfnamefont {M.}~\bibnamefont {Takigawa}}, \bibinfo {author} {\bibfnamefont {R.}~\bibnamefont {Higashinaka}}, \ and\ \bibinfo {author} {\bibfnamefont {Y.}~\bibnamefont {Maeno}},\ }\href@noop {} {\bibfield  {journal} {\bibinfo  {journal} {Physical review letters}\ }\textbf {\bibinfo {volume} {108}},\ \bibinfo {pages} {217203} (\bibinfo {year} {2012})}\BibitemShut {NoStop}%
\bibitem [{\citenamefont {Farhan}\ \emph {et~al.}(2013)\citenamefont {Farhan}, \citenamefont {Derlet}, \citenamefont {Kleibert}, \citenamefont {Balan}, \citenamefont {Chopdekar}, \citenamefont {Wyss}, \citenamefont {Perron}, \citenamefont {Scholl}, \citenamefont {Nolting},\ and\ \citenamefont {Heyderman}}]{Farhan2013}%
  \BibitemOpen
  \bibfield  {author} {\bibinfo {author} {\bibfnamefont {A.}~\bibnamefont {Farhan}}, \bibinfo {author} {\bibfnamefont {P.~M.}\ \bibnamefont {Derlet}}, \bibinfo {author} {\bibfnamefont {A.}~\bibnamefont {Kleibert}}, \bibinfo {author} {\bibfnamefont {A.}~\bibnamefont {Balan}}, \bibinfo {author} {\bibfnamefont {R.~V.}\ \bibnamefont {Chopdekar}}, \bibinfo {author} {\bibfnamefont {M.}~\bibnamefont {Wyss}}, \bibinfo {author} {\bibfnamefont {J.}~\bibnamefont {Perron}}, \bibinfo {author} {\bibfnamefont {A.}~\bibnamefont {Scholl}}, \bibinfo {author} {\bibfnamefont {F.}~\bibnamefont {Nolting}}, \ and\ \bibinfo {author} {\bibfnamefont {L.~J.}\ \bibnamefont {Heyderman}},\ }\href@noop {} {\bibfield  {journal} {\bibinfo  {journal} {Physical review letters}\ }\textbf {\bibinfo {volume} {111}},\ \bibinfo {pages} {057204} (\bibinfo {year} {2013})}\BibitemShut {NoStop}%
\bibitem [{\citenamefont {Pan}\ \emph {et~al.}(2016)\citenamefont {Pan}, \citenamefont {Laurita}, \citenamefont {Ross}, \citenamefont {Gaulin},\ and\ \citenamefont {Armitage}}]{Pan2016}%
  \BibitemOpen
  \bibfield  {author} {\bibinfo {author} {\bibfnamefont {L.}~\bibnamefont {Pan}}, \bibinfo {author} {\bibfnamefont {N.}~\bibnamefont {Laurita}}, \bibinfo {author} {\bibfnamefont {K.~A.}\ \bibnamefont {Ross}}, \bibinfo {author} {\bibfnamefont {B.~D.}\ \bibnamefont {Gaulin}}, \ and\ \bibinfo {author} {\bibfnamefont {N.}~\bibnamefont {Armitage}},\ }\href@noop {} {\bibfield  {journal} {\bibinfo  {journal} {Nature Physics}\ }\textbf {\bibinfo {volume} {12}},\ \bibinfo {pages} {361} (\bibinfo {year} {2016})}\BibitemShut {NoStop}%
\bibitem [{\citenamefont {Schiffer}\ and\ \citenamefont {Nisoli}(2021)}]{schiffer2021artificial}%
  \BibitemOpen
  \bibfield  {author} {\bibinfo {author} {\bibfnamefont {P.}~\bibnamefont {Schiffer}}\ and\ \bibinfo {author} {\bibfnamefont {C.}~\bibnamefont {Nisoli}},\ }\href@noop {} {\bibfield  {journal} {\bibinfo  {journal} {Applied Physics Letters}\ }\textbf {\bibinfo {volume} {118}} (\bibinfo {year} {2021})}\BibitemShut {NoStop}%
\bibitem [{\citenamefont {Harris}\ \emph {et~al.}(1997)\citenamefont {Harris}, \citenamefont {Bramwell}, \citenamefont {McMorrow}, \citenamefont {Zeiske},\ and\ \citenamefont {Godfrey}}]{Harris1997}%
  \BibitemOpen
  \bibfield  {author} {\bibinfo {author} {\bibfnamefont {M.~J.}\ \bibnamefont {Harris}}, \bibinfo {author} {\bibfnamefont {S.}~\bibnamefont {Bramwell}}, \bibinfo {author} {\bibfnamefont {D.}~\bibnamefont {McMorrow}}, \bibinfo {author} {\bibfnamefont {T.}~\bibnamefont {Zeiske}}, \ and\ \bibinfo {author} {\bibfnamefont {K.}~\bibnamefont {Godfrey}},\ }\href@noop {} {\bibfield  {journal} {\bibinfo  {journal} {Physical Review Letters}\ }\textbf {\bibinfo {volume} {79}},\ \bibinfo {pages} {2554} (\bibinfo {year} {1997})}\BibitemShut {NoStop}%
\bibitem [{\citenamefont {Fennell}\ \emph {et~al.}(2005)\citenamefont {Fennell}, \citenamefont {Petrenko}, \citenamefont {F{\aa}k}, \citenamefont {Gardner}, \citenamefont {Bramwell},\ and\ \citenamefont {Ouladdiaf}}]{fennell2005neutron}%
  \BibitemOpen
  \bibfield  {author} {\bibinfo {author} {\bibfnamefont {T.}~\bibnamefont {Fennell}}, \bibinfo {author} {\bibfnamefont {O.}~\bibnamefont {Petrenko}}, \bibinfo {author} {\bibfnamefont {B.}~\bibnamefont {F{\aa}k}}, \bibinfo {author} {\bibfnamefont {J.}~\bibnamefont {Gardner}}, \bibinfo {author} {\bibfnamefont {S.}~\bibnamefont {Bramwell}}, \ and\ \bibinfo {author} {\bibfnamefont {B.}~\bibnamefont {Ouladdiaf}},\ }\href@noop {} {\bibfield  {journal} {\bibinfo  {journal} {Physical Review B—Condensed Matter and Materials Physics}\ }\textbf {\bibinfo {volume} {72}},\ \bibinfo {pages} {224411} (\bibinfo {year} {2005})}\BibitemShut {NoStop}%
\bibitem [{\citenamefont {Jaubert}\ and\ \citenamefont {Holdsworth}(2011)}]{jaubert2011magnetic}%
  \BibitemOpen
  \bibfield  {author} {\bibinfo {author} {\bibfnamefont {L.~D.}\ \bibnamefont {Jaubert}}\ and\ \bibinfo {author} {\bibfnamefont {P.~C.}\ \bibnamefont {Holdsworth}},\ }\href@noop {} {\bibfield  {journal} {\bibinfo  {journal} {Journal of Physics: Condensed Matter}\ }\textbf {\bibinfo {volume} {23}},\ \bibinfo {pages} {164222} (\bibinfo {year} {2011})}\BibitemShut {NoStop}%
\bibitem [{\citenamefont {Fennell}\ \emph {et~al.}(2007)\citenamefont {Fennell}, \citenamefont {Bramwell}, \citenamefont {McMorrow}, \citenamefont {Manuel},\ and\ \citenamefont {Wildes}}]{Fennell2007}%
  \BibitemOpen
  \bibfield  {author} {\bibinfo {author} {\bibfnamefont {T.}~\bibnamefont {Fennell}}, \bibinfo {author} {\bibfnamefont {S.}~\bibnamefont {Bramwell}}, \bibinfo {author} {\bibfnamefont {D.}~\bibnamefont {McMorrow}}, \bibinfo {author} {\bibfnamefont {P.}~\bibnamefont {Manuel}}, \ and\ \bibinfo {author} {\bibfnamefont {A.}~\bibnamefont {Wildes}},\ }\href@noop {} {\bibfield  {journal} {\bibinfo  {journal} {Nature Physics}\ }\textbf {\bibinfo {volume} {3}},\ \bibinfo {pages} {566} (\bibinfo {year} {2007})}\BibitemShut {NoStop}%
\bibitem [{\citenamefont {Morris}\ \emph {et~al.}(2009)\citenamefont {Morris}, \citenamefont {Tennant}, \citenamefont {Grigera}, \citenamefont {Klemke}, \citenamefont {Castelnovo}, \citenamefont {Moessner}, \citenamefont {Czternasty}, \citenamefont {Meissner}, \citenamefont {Rule}, \citenamefont {Hoffmann} \emph {et~al.}}]{Morris2009}%
  \BibitemOpen
  \bibfield  {author} {\bibinfo {author} {\bibfnamefont {D.~J.~P.}\ \bibnamefont {Morris}}, \bibinfo {author} {\bibfnamefont {D.~A.}\ \bibnamefont {Tennant}}, \bibinfo {author} {\bibfnamefont {S.~A.}\ \bibnamefont {Grigera}}, \bibinfo {author} {\bibfnamefont {B.}~\bibnamefont {Klemke}}, \bibinfo {author} {\bibfnamefont {C.}~\bibnamefont {Castelnovo}}, \bibinfo {author} {\bibfnamefont {R.}~\bibnamefont {Moessner}}, \bibinfo {author} {\bibfnamefont {C.}~\bibnamefont {Czternasty}}, \bibinfo {author} {\bibfnamefont {M.}~\bibnamefont {Meissner}}, \bibinfo {author} {\bibfnamefont {K.}~\bibnamefont {Rule}}, \bibinfo {author} {\bibfnamefont {J.-U.}\ \bibnamefont {Hoffmann}},  \emph {et~al.},\ }\href@noop {} {\bibfield  {journal} {\bibinfo  {journal} {Science}\ }\textbf {\bibinfo {volume} {326}},\ \bibinfo {pages} {411} (\bibinfo {year} {2009})}\BibitemShut {NoStop}%
\bibitem [{\citenamefont {Sarkar}\ and\ \citenamefont {Mukhopadhyay}(2014)}]{sarkar2014dynamics}%
  \BibitemOpen
  \bibfield  {author} {\bibinfo {author} {\bibfnamefont {A.}~\bibnamefont {Sarkar}}\ and\ \bibinfo {author} {\bibfnamefont {S.}~\bibnamefont {Mukhopadhyay}},\ }\href@noop {} {\bibfield  {journal} {\bibinfo  {journal} {Physical Review B}\ }\textbf {\bibinfo {volume} {90}},\ \bibinfo {pages} {165129} (\bibinfo {year} {2014})}\BibitemShut {NoStop}%
\bibitem [{\citenamefont {Edberg}\ \emph {et~al.}(2025)\citenamefont {Edberg}, \citenamefont {Khansili}, \citenamefont {Fjellv\aa{}g}, \citenamefont {\O{}rduk~Sandberg}, \citenamefont {Deen}, \citenamefont {Lefmann}, \citenamefont {Henelius},\ and\ \citenamefont {Rydh}}]{Edberg2025}%
  \BibitemOpen
  \bibfield  {author} {\bibinfo {author} {\bibfnamefont {R.}~\bibnamefont {Edberg}}, \bibinfo {author} {\bibfnamefont {A.}~\bibnamefont {Khansili}}, \bibinfo {author} {\bibfnamefont {I.~M.}\ \bibnamefont {Fjellv\aa{}g}}, \bibinfo {author} {\bibfnamefont {L.}~\bibnamefont {\O{}rduk~Sandberg}}, \bibinfo {author} {\bibfnamefont {P.~P.}\ \bibnamefont {Deen}}, \bibinfo {author} {\bibfnamefont {K.}~\bibnamefont {Lefmann}}, \bibinfo {author} {\bibfnamefont {P.}~\bibnamefont {Henelius}}, \ and\ \bibinfo {author} {\bibfnamefont {A.}~\bibnamefont {Rydh}},\ }\href {\doibase 10.1103/PhysRevB.112.094435} {\bibfield  {journal} {\bibinfo  {journal} {Phys. Rev. B}\ }\textbf {\bibinfo {volume} {112}},\ \bibinfo {pages} {094435} (\bibinfo {year} {2025})}\BibitemShut {NoStop}%
\bibitem [{\citenamefont {Bramwell}\ and\ \citenamefont {Gingras}(2001)}]{Bramwell2001}%
  \BibitemOpen
  \bibfield  {author} {\bibinfo {author} {\bibfnamefont {S.~T.}\ \bibnamefont {Bramwell}}\ and\ \bibinfo {author} {\bibfnamefont {M.~J.}\ \bibnamefont {Gingras}},\ }\href@noop {} {\bibfield  {journal} {\bibinfo  {journal} {Science}\ }\textbf {\bibinfo {volume} {294}},\ \bibinfo {pages} {1495} (\bibinfo {year} {2001})}\BibitemShut {NoStop}%
\bibitem [{\citenamefont {Snyder}\ \emph {et~al.}(2004)\citenamefont {Snyder}, \citenamefont {Ueland}, \citenamefont {Slusky}, \citenamefont {Karunadasa}, \citenamefont {Cava},\ and\ \citenamefont {Schiffer}}]{snyder2004low}%
  \BibitemOpen
  \bibfield  {author} {\bibinfo {author} {\bibfnamefont {J.}~\bibnamefont {Snyder}}, \bibinfo {author} {\bibfnamefont {B.}~\bibnamefont {Ueland}}, \bibinfo {author} {\bibfnamefont {J.}~\bibnamefont {Slusky}}, \bibinfo {author} {\bibfnamefont {H.}~\bibnamefont {Karunadasa}}, \bibinfo {author} {\bibfnamefont {R.}~\bibnamefont {Cava}}, \ and\ \bibinfo {author} {\bibfnamefont {P.}~\bibnamefont {Schiffer}},\ }\href@noop {} {\bibfield  {journal} {\bibinfo  {journal} {Physical Review B}\ }\textbf {\bibinfo {volume} {69}},\ \bibinfo {pages} {064414} (\bibinfo {year} {2004})}\BibitemShut {NoStop}%
\bibitem [{\citenamefont {Powell}\ and\ \citenamefont {Pal}(2025)}]{powell2025dynamic}%
  \BibitemOpen
  \bibfield  {author} {\bibinfo {author} {\bibfnamefont {S.}~\bibnamefont {Powell}}\ and\ \bibinfo {author} {\bibfnamefont {S.}~\bibnamefont {Pal}},\ }\href@noop {} {\bibfield  {journal} {\bibinfo  {journal} {Physical Review Letters}\ }\textbf {\bibinfo {volume} {134}},\ \bibinfo {pages} {256701} (\bibinfo {year} {2025})}\BibitemShut {NoStop}%
\bibitem [{\citenamefont {Dusad}\ \emph {et~al.}(2019)\citenamefont {Dusad}, \citenamefont {Kirschner}, \citenamefont {Hoke}, \citenamefont {Roberts}, \citenamefont {Eyal}, \citenamefont {Flicker}, \citenamefont {Luke}, \citenamefont {Blundell},\ and\ \citenamefont {Davis}}]{Dusad2019}%
  \BibitemOpen
  \bibfield  {author} {\bibinfo {author} {\bibfnamefont {R.}~\bibnamefont {Dusad}}, \bibinfo {author} {\bibfnamefont {F.~K.}\ \bibnamefont {Kirschner}}, \bibinfo {author} {\bibfnamefont {J.~C.}\ \bibnamefont {Hoke}}, \bibinfo {author} {\bibfnamefont {B.~R.}\ \bibnamefont {Roberts}}, \bibinfo {author} {\bibfnamefont {A.}~\bibnamefont {Eyal}}, \bibinfo {author} {\bibfnamefont {F.}~\bibnamefont {Flicker}}, \bibinfo {author} {\bibfnamefont {G.~M.}\ \bibnamefont {Luke}}, \bibinfo {author} {\bibfnamefont {S.~J.}\ \bibnamefont {Blundell}}, \ and\ \bibinfo {author} {\bibfnamefont {J.~S.}\ \bibnamefont {Davis}},\ }\href@noop {} {\bibfield  {journal} {\bibinfo  {journal} {Nature}\ }\textbf {\bibinfo {volume} {571}},\ \bibinfo {pages} {234} (\bibinfo {year} {2019})}\BibitemShut {NoStop}%
\bibitem [{\citenamefont {Miao}\ \emph {et~al.}(2020)\citenamefont {Miao}, \citenamefont {Lee}, \citenamefont {Mei}, \citenamefont {Lawler},\ and\ \citenamefont {Shen}}]{Miao2020}%
  \BibitemOpen
  \bibfield  {author} {\bibinfo {author} {\bibfnamefont {L.}~\bibnamefont {Miao}}, \bibinfo {author} {\bibfnamefont {Y.}~\bibnamefont {Lee}}, \bibinfo {author} {\bibfnamefont {A.}~\bibnamefont {Mei}}, \bibinfo {author} {\bibfnamefont {M.}~\bibnamefont {Lawler}}, \ and\ \bibinfo {author} {\bibfnamefont {K.}~\bibnamefont {Shen}},\ }\href@noop {} {\bibfield  {journal} {\bibinfo  {journal} {Nature Communications}\ }\textbf {\bibinfo {volume} {11}},\ \bibinfo {pages} {1341} (\bibinfo {year} {2020})}\BibitemShut {NoStop}%
\bibitem [{\citenamefont {Timsina}\ \emph {et~al.}(2024)\citenamefont {Timsina}, \citenamefont {Kiefer},\ and\ \citenamefont {Miao}}]{Timsina2024}%
  \BibitemOpen
  \bibfield  {author} {\bibinfo {author} {\bibfnamefont {P.}~\bibnamefont {Timsina}}, \bibinfo {author} {\bibfnamefont {B.}~\bibnamefont {Kiefer}}, \ and\ \bibinfo {author} {\bibfnamefont {L.}~\bibnamefont {Miao}},\ }\href@noop {} {\bibfield  {journal} {\bibinfo  {journal} {Physical Review B}\ }\textbf {\bibinfo {volume} {110}},\ \bibinfo {pages} {184420} (\bibinfo {year} {2024})}\BibitemShut {NoStop}%
\bibitem [{\citenamefont {Lefran{\c{c}}ois}\ \emph {et~al.}(2017)\citenamefont {Lefran{\c{c}}ois}, \citenamefont {Cathelin}, \citenamefont {Lhotel}, \citenamefont {Robert}, \citenamefont {Lejay}, \citenamefont {Colin}, \citenamefont {Canals}, \citenamefont {Damay}, \citenamefont {Ollivier}, \citenamefont {F{\aa}k} \emph {et~al.}}]{Lefrancois2017}%
  \BibitemOpen
  \bibfield  {author} {\bibinfo {author} {\bibfnamefont {E.}~\bibnamefont {Lefran{\c{c}}ois}}, \bibinfo {author} {\bibfnamefont {V.}~\bibnamefont {Cathelin}}, \bibinfo {author} {\bibfnamefont {E.}~\bibnamefont {Lhotel}}, \bibinfo {author} {\bibfnamefont {J.}~\bibnamefont {Robert}}, \bibinfo {author} {\bibfnamefont {P.}~\bibnamefont {Lejay}}, \bibinfo {author} {\bibfnamefont {C.~V.}\ \bibnamefont {Colin}}, \bibinfo {author} {\bibfnamefont {B.}~\bibnamefont {Canals}}, \bibinfo {author} {\bibfnamefont {F.}~\bibnamefont {Damay}}, \bibinfo {author} {\bibfnamefont {J.}~\bibnamefont {Ollivier}}, \bibinfo {author} {\bibfnamefont {B.}~\bibnamefont {F{\aa}k}},  \emph {et~al.},\ }\href@noop {} {\bibfield  {journal} {\bibinfo  {journal} {Nature communications}\ }\textbf {\bibinfo {volume} {8}},\ \bibinfo {pages} {209} (\bibinfo {year} {2017})}\BibitemShut {NoStop}%
\bibitem [{\citenamefont {Timsina}\ \emph {et~al.}(2025)\citenamefont {Timsina}, \citenamefont {Chappa}, \citenamefont {Alyones}, \citenamefont {Vasiliev},\ and\ \citenamefont {Miao}}]{timsina2025emergent}%
  \BibitemOpen
  \bibfield  {author} {\bibinfo {author} {\bibfnamefont {P.}~\bibnamefont {Timsina}}, \bibinfo {author} {\bibfnamefont {A.}~\bibnamefont {Chappa}}, \bibinfo {author} {\bibfnamefont {D.}~\bibnamefont {Alyones}}, \bibinfo {author} {\bibfnamefont {I.}~\bibnamefont {Vasiliev}}, \ and\ \bibinfo {author} {\bibfnamefont {L.}~\bibnamefont {Miao}},\ }\href@noop {} {\bibfield  {journal} {\bibinfo  {journal} {arXiv preprint arXiv:2505.13352}\ } (\bibinfo {year} {2025})}\BibitemShut {NoStop}%
\bibitem [{\citenamefont {Guo}\ \emph {et~al.}(2016)\citenamefont {Guo}, \citenamefont {Ritter},\ and\ \citenamefont {Komarek}}]{Guo2016}%
  \BibitemOpen
  \bibfield  {author} {\bibinfo {author} {\bibfnamefont {H.}~\bibnamefont {Guo}}, \bibinfo {author} {\bibfnamefont {C.}~\bibnamefont {Ritter}}, \ and\ \bibinfo {author} {\bibfnamefont {A.}~\bibnamefont {Komarek}},\ }\href@noop {} {\bibfield  {journal} {\bibinfo  {journal} {Physical Review B}\ }\textbf {\bibinfo {volume} {94}},\ \bibinfo {pages} {161102} (\bibinfo {year} {2016})}\BibitemShut {NoStop}%
\bibitem [{\citenamefont {Onoda}\ and\ \citenamefont {Ishii}(2019)}]{Onoda2019}%
  \BibitemOpen
  \bibfield  {author} {\bibinfo {author} {\bibfnamefont {S.}~\bibnamefont {Onoda}}\ and\ \bibinfo {author} {\bibfnamefont {F.}~\bibnamefont {Ishii}},\ }\href@noop {} {\bibfield  {journal} {\bibinfo  {journal} {Physical review letters}\ }\textbf {\bibinfo {volume} {122}},\ \bibinfo {pages} {067201} (\bibinfo {year} {2019})}\BibitemShut {NoStop}%
\bibitem [{\citenamefont {Nagaosa}\ and\ \citenamefont {Tokura}(2013)}]{nagaosa2013topological}%
  \BibitemOpen
  \bibfield  {author} {\bibinfo {author} {\bibfnamefont {N.}~\bibnamefont {Nagaosa}}\ and\ \bibinfo {author} {\bibfnamefont {Y.}~\bibnamefont {Tokura}},\ }\href@noop {} {\bibfield  {journal} {\bibinfo  {journal} {Nature nanotechnology}\ }\textbf {\bibinfo {volume} {8}},\ \bibinfo {pages} {899} (\bibinfo {year} {2013})}\BibitemShut {NoStop}%
\bibitem [{\citenamefont {Fert}\ \emph {et~al.}(2017)\citenamefont {Fert}, \citenamefont {Reyren},\ and\ \citenamefont {Cros}}]{fert2017magnetic}%
  \BibitemOpen
  \bibfield  {author} {\bibinfo {author} {\bibfnamefont {A.}~\bibnamefont {Fert}}, \bibinfo {author} {\bibfnamefont {N.}~\bibnamefont {Reyren}}, \ and\ \bibinfo {author} {\bibfnamefont {V.}~\bibnamefont {Cros}},\ }\href@noop {} {\bibfield  {journal} {\bibinfo  {journal} {Nature Reviews Materials}\ }\textbf {\bibinfo {volume} {2}},\ \bibinfo {pages} {1} (\bibinfo {year} {2017})}\BibitemShut {NoStop}%
\bibitem [{\citenamefont {Li}\ \emph {et~al.}(2023)\citenamefont {Li}, \citenamefont {Wang},\ and\ \citenamefont {Rasing}}]{li2023magnetic}%
  \BibitemOpen
  \bibfield  {author} {\bibinfo {author} {\bibfnamefont {S.}~\bibnamefont {Li}}, \bibinfo {author} {\bibfnamefont {X.}~\bibnamefont {Wang}}, \ and\ \bibinfo {author} {\bibfnamefont {T.}~\bibnamefont {Rasing}},\ }\href@noop {} {\bibfield  {journal} {\bibinfo  {journal} {Interdisciplinary Materials}\ }\textbf {\bibinfo {volume} {2}},\ \bibinfo {pages} {260} (\bibinfo {year} {2023})}\BibitemShut {NoStop}%
\bibitem [{\citenamefont {Pan}\ \emph {et~al.}(2014)\citenamefont {Pan}, \citenamefont {Kim}, \citenamefont {Ghosh}, \citenamefont {Morris}, \citenamefont {Ross}, \citenamefont {Kermarrec}, \citenamefont {Gaulin}, \citenamefont {Koohpayeh}, \citenamefont {Tchernyshyov},\ and\ \citenamefont {Armitage}}]{Pan2014}%
  \BibitemOpen
  \bibfield  {author} {\bibinfo {author} {\bibfnamefont {L.}~\bibnamefont {Pan}}, \bibinfo {author} {\bibfnamefont {S.~K.}\ \bibnamefont {Kim}}, \bibinfo {author} {\bibfnamefont {A.}~\bibnamefont {Ghosh}}, \bibinfo {author} {\bibfnamefont {C.~M.}\ \bibnamefont {Morris}}, \bibinfo {author} {\bibfnamefont {K.~A.}\ \bibnamefont {Ross}}, \bibinfo {author} {\bibfnamefont {E.}~\bibnamefont {Kermarrec}}, \bibinfo {author} {\bibfnamefont {B.~D.}\ \bibnamefont {Gaulin}}, \bibinfo {author} {\bibfnamefont {S.}~\bibnamefont {Koohpayeh}}, \bibinfo {author} {\bibfnamefont {O.}~\bibnamefont {Tchernyshyov}}, \ and\ \bibinfo {author} {\bibfnamefont {N.}~\bibnamefont {Armitage}},\ }\href@noop {} {\bibfield  {journal} {\bibinfo  {journal} {Nature communications}\ }\textbf {\bibinfo {volume} {5}},\ \bibinfo {pages} {1} (\bibinfo {year} {2014})}\BibitemShut {NoStop}%
\end{thebibliography}%

\section*{Acknowledgements} This work was supported by the National Science Foundation's Partnerships for Research and Education in Materials with award number 2423992.
\section*{Author Contribution} L.M. designed and directed the research. P.T. created and executed the simulation programs. The data were analyzed by P.T., A.C., and D.A., and interpreted by B.K.

\section*{Additional information}
Correspondence and requests for materials should be addressed to L.M.
\section*{Competing financial interests}
The authors declare no competing interests.

\end{document}